\documentclass[12pt]{article}
\usepackage{graphics}
\usepackage{a4wide}
\usepackage{cite}   

\newcommand{\GeV}{\mbox{ GeV}}
\newcommand{\TeV}{\mbox{ TeV}}

\newcommand{\sq}{\ensuremath{\tilde q}}
\newcommand{\stopp}{\ensuremath{\tilde t}}
\newcommand{\sbottom}{\ensuremath{\tilde b}} 

\newcommand{\cmin}{\ensuremath{\chi^-}}
\newcommand{\neut}{\ensuremath{\chi^0}}

\newcommand{\PL}{\ensuremath{P_L}}
\newcommand{\PR}{\ensuremath{P_R}}

\newcommand{\slas}[1]{\rlap/ #1}

\newcommand{\lsim}{\mbox{ \raisebox{-4pt}{${\stackrel{\textstyle <}{\sim}}$} }}

\newcommand{\sw}{\ensuremath{\sin \theta_W}}
\newcommand{\tb}{\ensuremath{\tan\beta}}

\newcommand{\mb}{\ensuremath{m_b}}
\newcommand{\mbs}{\ensuremath{m_b^2}}
\newcommand{\mt}{\ensuremath{m_t}}
\newcommand{\lb}{\ensuremath{\lambda_b}}
\newcommand{\mw}{\ensuremath{M_W}}
\newcommand{\mws}{\ensuremath{M^2_W}}

\newcommand{\mzs}{\ensuremath{M^2_Z}}
\newcommand{\mHp}{\ensuremath{M_{H^\pm}}}
\newcommand{\msbas}{\ensuremath{m_{\tilde{b}_a}^2}}

\newcommand{\sba}{\ensuremath{\tilde{b}_a}}
\newcommand{\nba}{\ensuremath{\bar{\Psi}_\alpha^0}}

\newcommand{\Robc}{\ensuremath{R_{1a}^{(b)}}}
\newcommand{\Rtbc}{\ensuremath{R_{2a}^{(b)}}}

\newcommand{\No}{\ensuremath{N_{\alpha1}}}
\newcommand{\Nt}{\ensuremath{N_{\alpha2}}}
\newcommand{\Nth}{\ensuremath{N_{\alpha3}}}

\newcommand{\Apab}{\ensuremath{A_{+a\alpha}^{(b)}}}
\newcommand{\Amab}{\ensuremath{A_{-a\alpha}^{(b)}}}

\newcommand{\Amabc}{\ensuremath{A_{-a\alpha}^{(b)*}}}

\newcommand{\Apmbbc}{\ensuremath{A_{\pm a\beta}^{(b)*}}}

\newcommand{\LApab}{\ensuremath{\Lambda_{+a\alpha}^{(b)}}}
\newcommand{\LAmab}{\ensuremath{\Lambda_{-a\alpha}^{(b)}}}

\newcommand{\Cpab}{\ensuremath{C_{+a\alpha}^{(b)}}}
\newcommand{\Cmab}{\ensuremath{C_{-a\alpha}^{(b)}}}

\newcommand{\LApmab}{\ensuremath{\Lambda_{\pm a\alpha}^{(b)}}}
\newcommand{\LApmabc}{\ensuremath{\Lambda_{\pm a\alpha}^{(b)*}}}

\newcommand{\dalpha}{\frac{\delta\alpha}{\alpha}}
\newcommand{\dmws}{\frac{\delta \mws}{\mws}}
\newcommand{\dmzs}{\frac{\delta \mzs}{\mzs}}
\newcommand{\dz}{\delta Z}
\newcommand{\dmb}{\frac{\delta \mb}{\mb}}
\newcommand{\Dcosb}{\frac{\delta \cos \beta}{\cos\beta}}
\newcommand{\DZlbot}{\delta Z_L^b}
\newcommand{\DZrbot}{\delta Z_R^b}
\newcommand{\Dzrneut}{\delta Z_R^0}
\newcommand{\Dzlneut}{\delta Z_L^0}
\newcommand{\Drb}{\delta R^{(b)}}
\newcommand{\swp}{s_W}
\newcommand{\cwp}{c_W}
\newcommand{\twp}{t_W}
\newcommand{\dtow}{\frac{\delta \twp}{\twp}}

\newcommand{\figsbottomone}{
\begin{figure}[t]
  \begin{center}
\begin{tabular}{cc}
\multicolumn{2}{c}{{\resizebox{7.2cm}{!}{\includegraphics{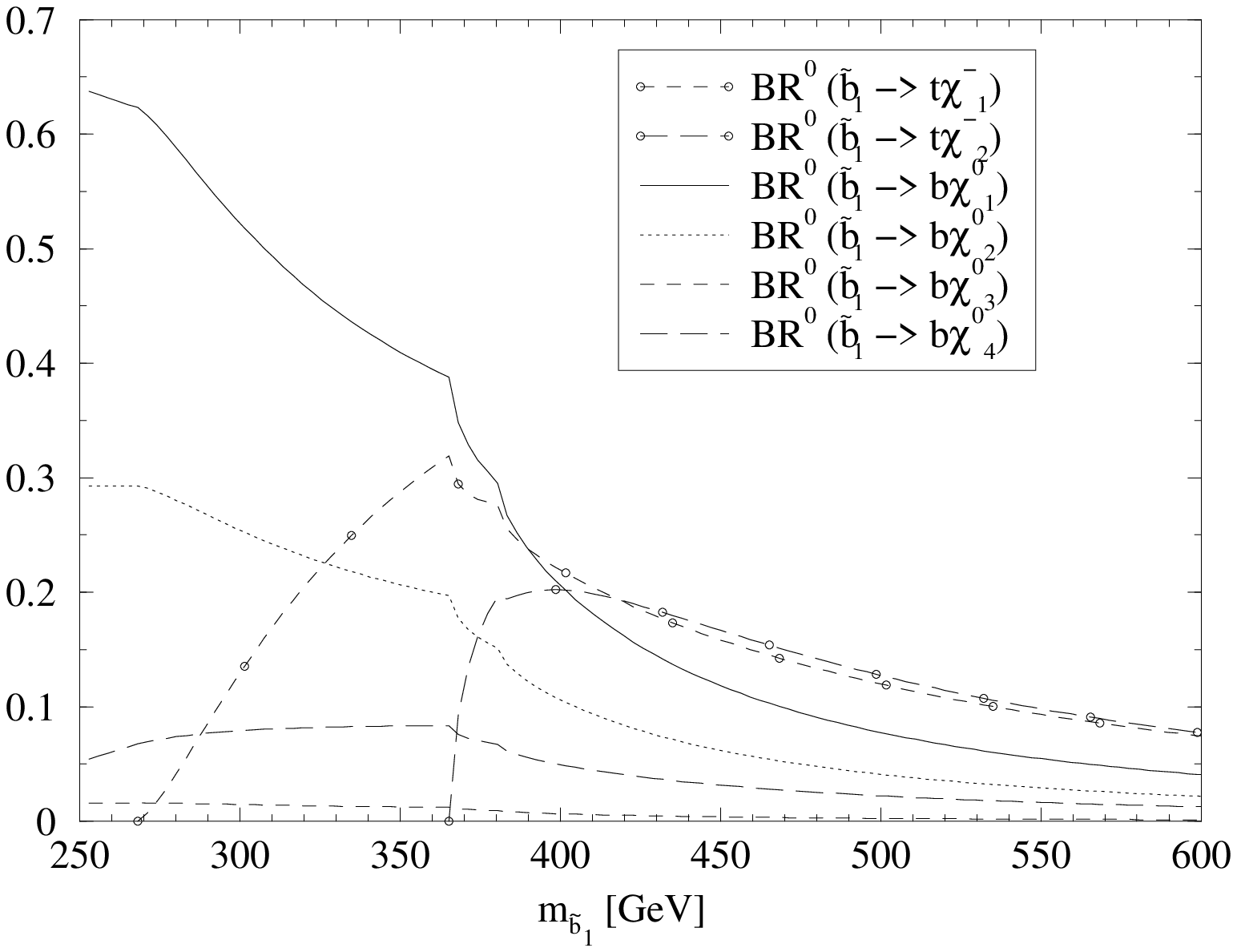}}}}\\
\multicolumn{2}{c}{(a)}\\
{\resizebox{7.2cm}{!}{\includegraphics{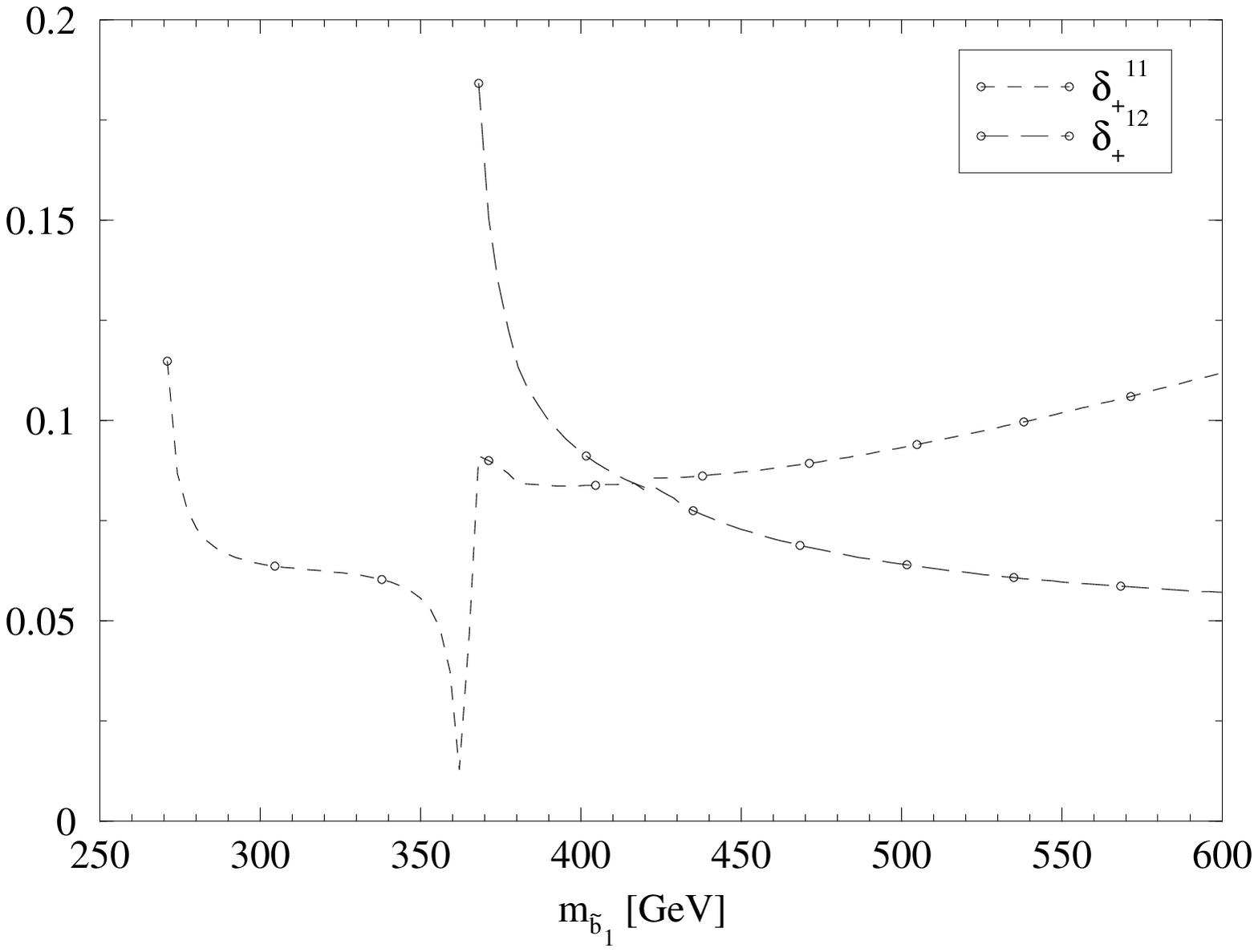}}} &
{\resizebox{7.2cm}{!}{\includegraphics{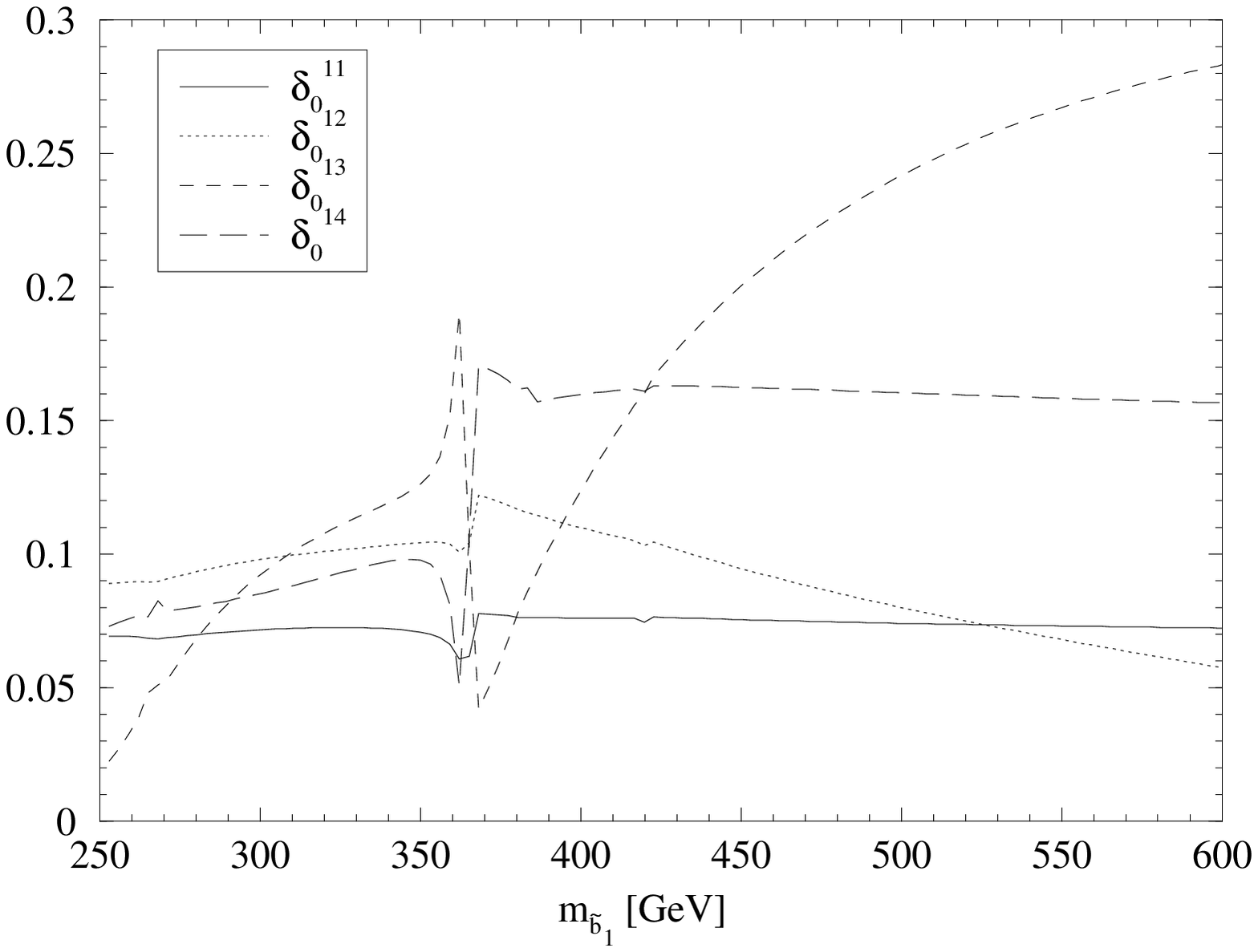}}}\\
(b) & (c)
\end{tabular}
    \caption{\textbf{(a)} Tree-level branching ratios for the lightest
      sbottom branching ratios into charginos and 
      neutralinos as a function of the lightest sbottom mass; \textbf{(b)} EW
      radiative corrections, eq.~(\ref{eq:deltadef}), to the partial widths of lightest sbottom
      decaying into charginos; \textbf{(c)} as in (b), but for the
      neutralino channels.} 
    \label{fig:sbottom}
  \end{center}
\end{figure}}

\newcommand{\figsbottomtwo}
{\begin{figure}[t]
  \begin{center}
\begin{tabular}{cc}
\multicolumn{2}{c}{\resizebox{7.5cm}{!}{\includegraphics{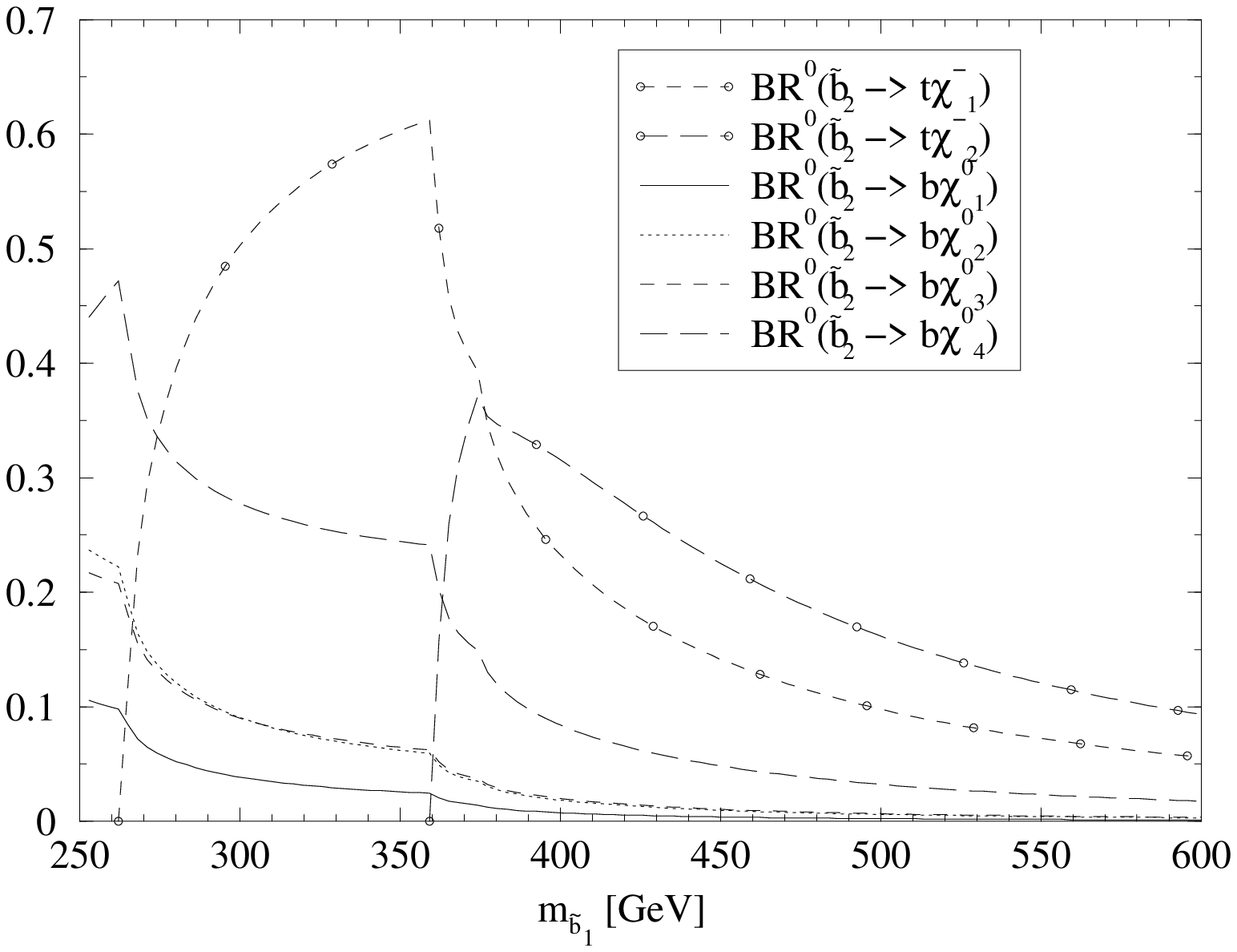}}}\\
\multicolumn{2}{c}{(a)}\\
\resizebox{7.5cm}{!}{\includegraphics{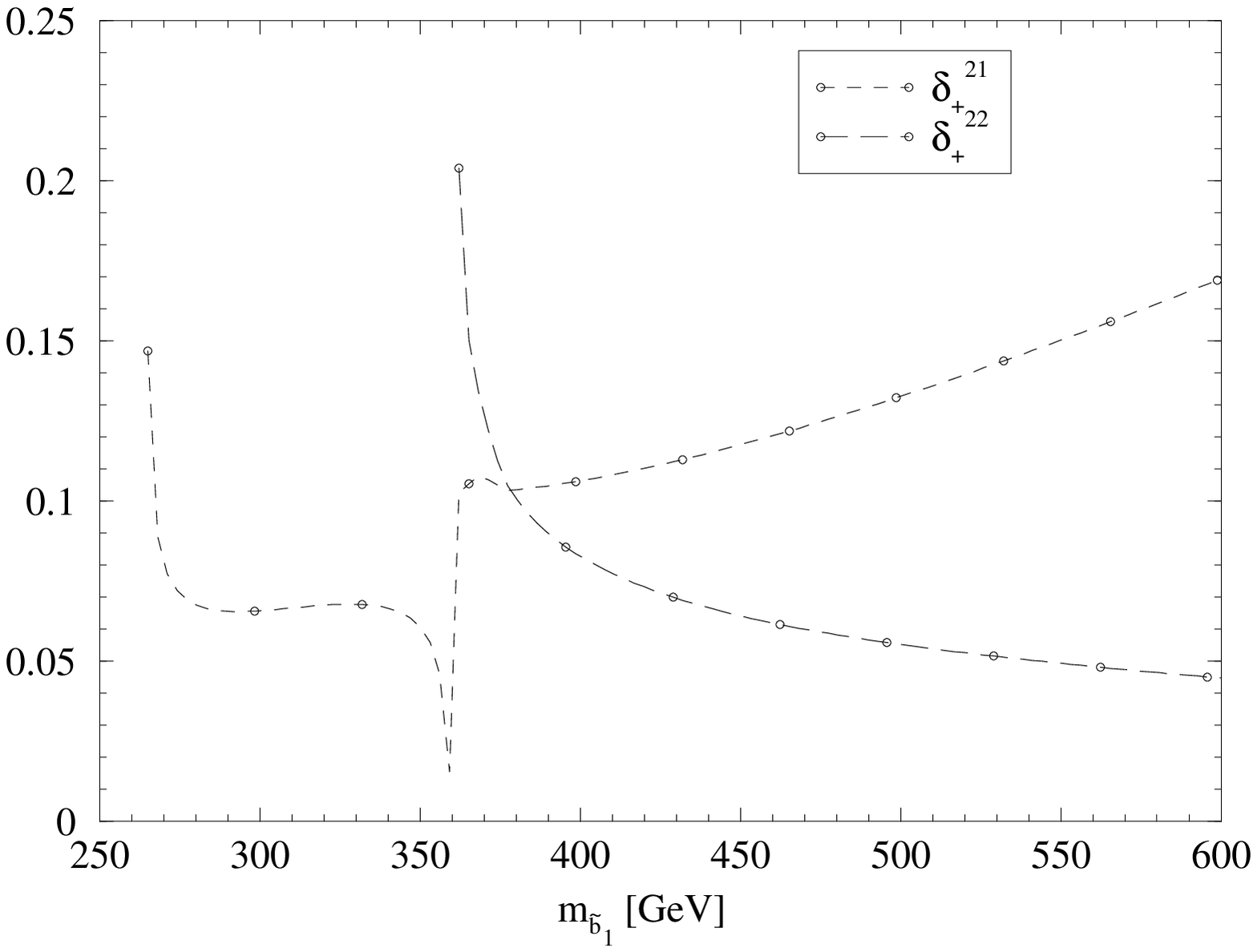}} &
\resizebox{7.5cm}{!}{\includegraphics{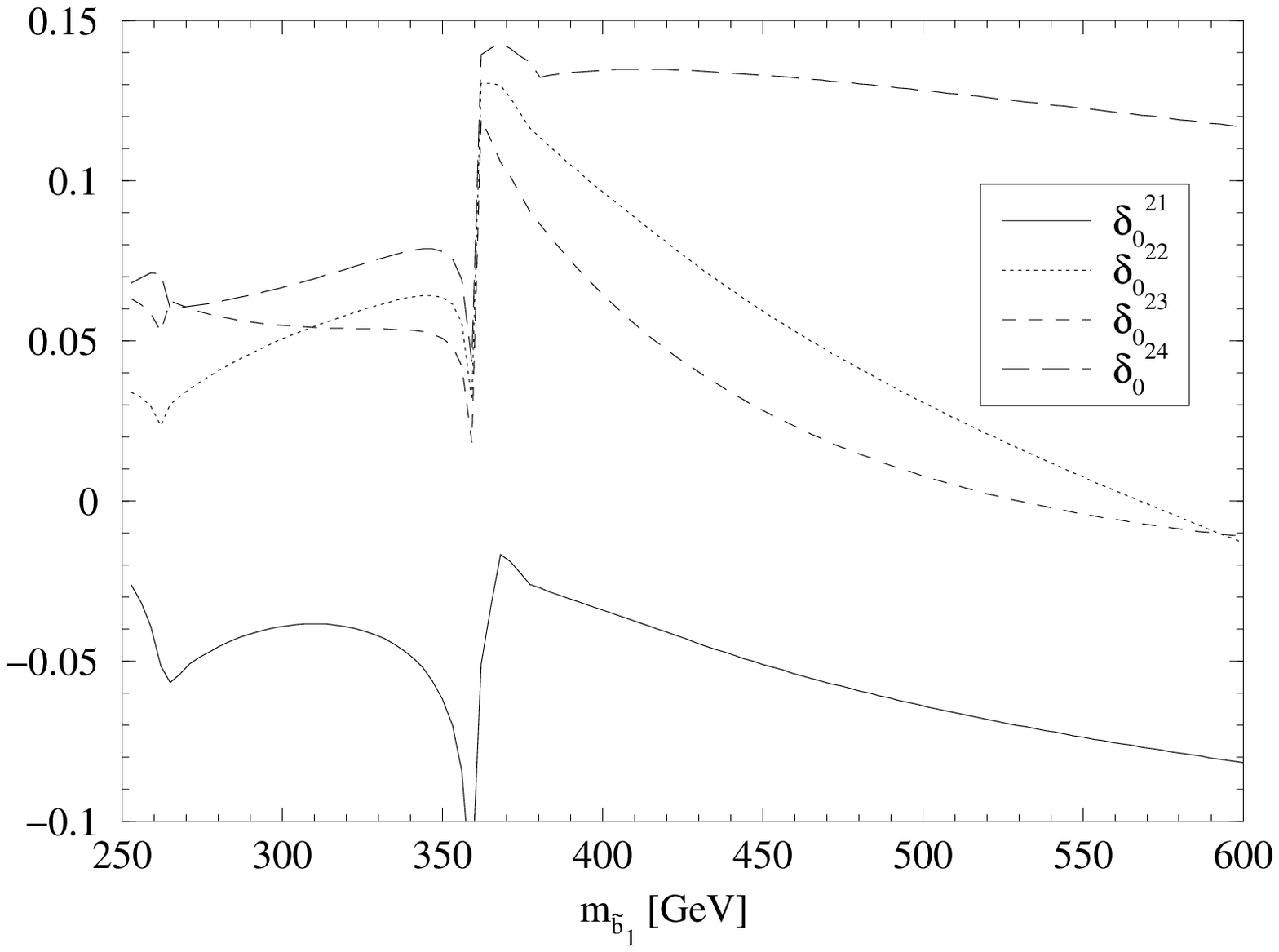}}\\
(b) & (c)
\end{tabular}
    \caption{As in Fig.~\ref{fig:sbottom} but for the heaviest sbottom.} 
    \label{fig:sbottom2}
  \end{center}
\end{figure}}

\newcommand{\figstop}
{\begin{figure}[t]
  \begin{center}
    \begin{tabular}{cc}
\resizebox{7.5cm}{!}{\includegraphics{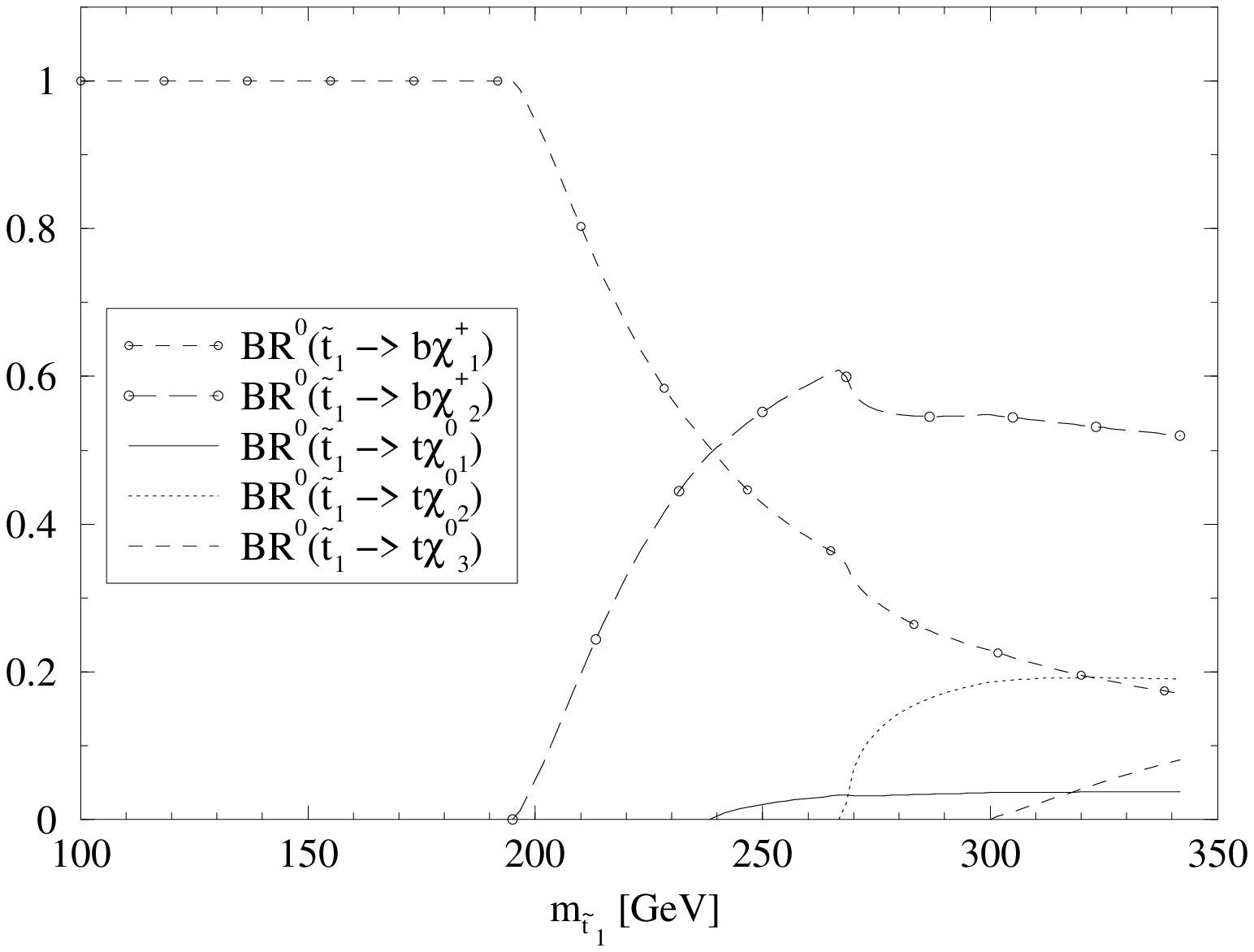}} &
\resizebox{7.5cm}{!}{\includegraphics{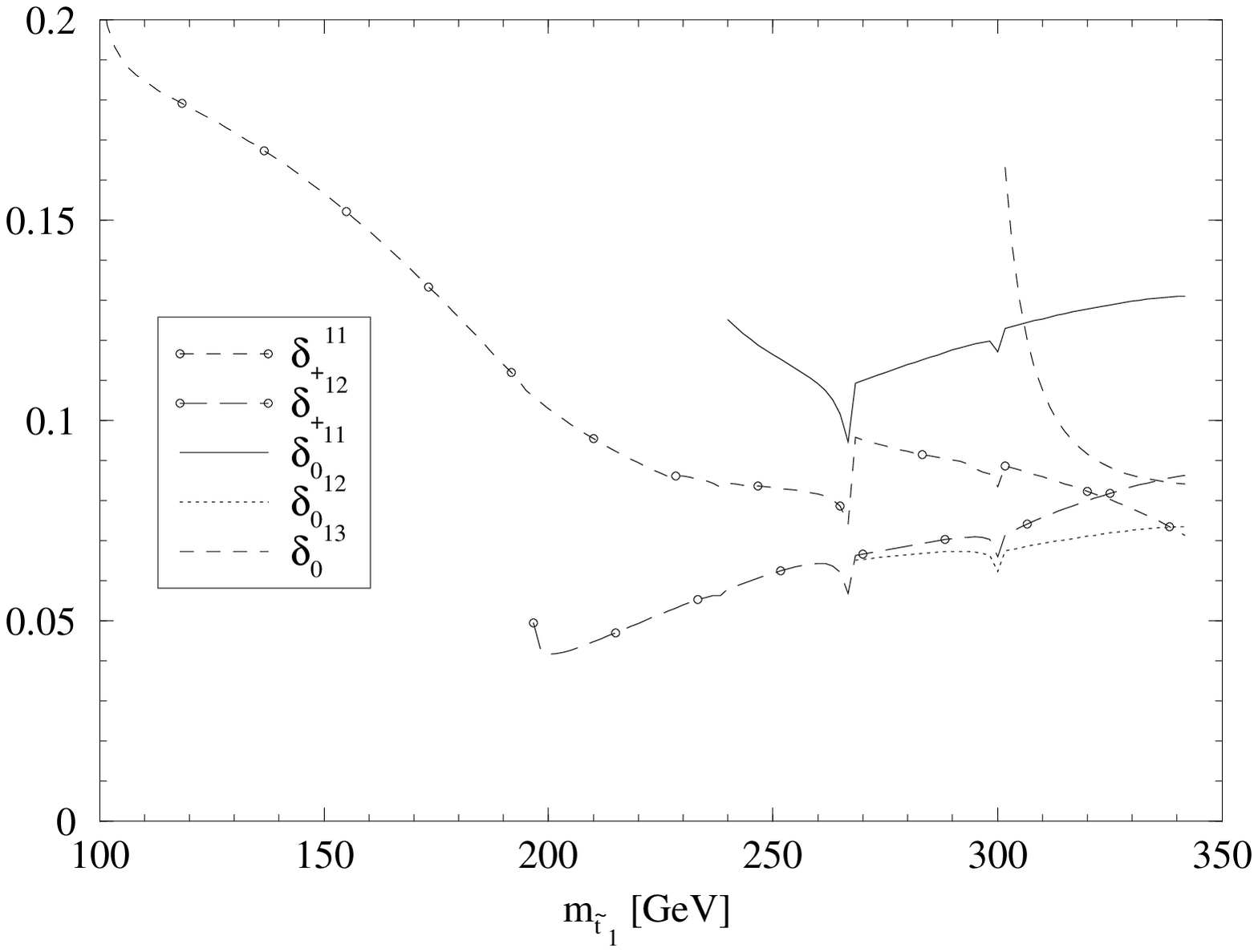}}\\
(b) & (c)
    \end{tabular}
    \caption{\textbf{(a)} Tree-level branching ratios for the lightest
      stop branching ratios into charginos and 
      neutralinos as a function of the lightest stop mass; \textbf{(b)} EW
      radiative corrections, eq.~(\ref{eq:deltadef}), to the partial
      widths of lightest stop 
      decaying into charginos and neutralinos.}
    \label{fig:stop}
  \end{center}
\end{figure}}

\begin{document}

\thispagestyle{empty}

\begin{flushright}
{\parbox{3.5cm}{KA-TP-25-2000\\
UAB-FT-505\\
hep-ph/0101086
}}
\end{flushright}

\vspace{0.5cm}

\begin{center}

{\Large \textbf{Full electroweak one-loop radiative corrections to
    squark decays in the MSSM}} 

\vspace{0.8cm}

{\large J.~Guasch$^{a}$, W.~Hollik$^{a}$, J.~Sol\`a$^{b}$}

\vspace*{0.8cm}

{\sl $^a$ Institut f\"ur Theoretische Physik, Universit\"at Karlsruhe,  D-76128 Karlsruhe, Germany\\
$^b$  Grup de F{\'\i}sica Te{\`o}rica and Institut de F{\'\i}sica d'Altes
Energies, Universitat Aut{\`o}noma de Barcelona, E-08193 Bellaterra
(Barcelona), Catalonia, Spain} 

\end{center}

\vspace*{0.5cm}

\begin{abstract} 
We present results on the full one-loop electroweak radiative
corrections to the squark decay partial widths into charginos and
neutralinos. We show the renormalization framework, and present
numerical results for the third squark family. The corrections can
reach values of $\sim 10\%$, which are comparable to the 
radiative corrections from the strong sector of the model. Therefore
they should be taken into account for the precise extraction of the SUSY
parameters at future colliders.
\end{abstract}
\newpage

The Standard Model (SM) of the strong and electroweak interactions is the
present paradigm of particle physics. Its validity has been tested to a
level better than one per mil in the particle
accelerators~\cite{PDB2000}. 
Nevertheless, there are
arguments against the SM being the fundamental model of
particle interactions~\cite{Carena:2000yx}, giving rise to the
investigation of competing alternative or extended models, which
can be tested at high-energy colliders, such as the Large Hadron Collider
(LHC)~\cite{ATLASCMS}, or a $e^+e^-$ Linear
Collider (LC)~\cite{Miller:1998fh}. 
One of the most promising possibilities for physics beyond
the SM is the incorporation of Supersymmetry (SUSY), which leads to
a  renormalizable field theory with precisely calculable
predictions to be tested in future experiments. 
The simplest supersymmetric
extension of the SM is the Minimal Supersymmetric
Standard Model
(MSSM)~\cite{MSSM}. 
If the masses of the extra non-standard particles
are very large as compared to the SM 
electroweak scale, the effects of
these particles decouple from the SM  low-energy effective
Lagrangian~\cite{Dobado:1997up}. 
This means that if the extra particles are
too heavy we could not discern between the SM and the MSSM by just
looking at the low-energy end of the spectrum, since the only trace of
the MSSM would be a light Higgs boson 
($M_{h^0}\lsim
135\GeV$)~\cite{Higgs2L}, 
whose properties would not differ from the SM one. 
Nevertheless, when some of the extra particle masses are of the order of
the electroweak scale,
the next generation of colliders will be able to
produce such kind of particles and investigate their properties. 
In this case non-decoupling effects appear~\cite{Djouadi:1997wt,GHS2}.
While the LHC will be able to produce
new  particles with masses up to $2.5\TeV$, the LC will be able to
make precision measurements of their properties, provided 
they are not too heavy.
For an adequate analysis, precise theoretical predictions are
required, going beyond the Born approximation for SUSY processes.

Up to now the major effort on the computation of SUSY radiative
corrections has been put into the computation of virtual SUSY effects in
observables that involve only SM external particles, 
or into the calculation of loop effects in the extended
Higgs sector of the MSSM\footnote{See e.g.~\cite{IWQEMSSM}
and references therein.}. 
But for the case of direct production of SUSY particles, one also
needs a detailed knowledge of the higher-order effects for the processes
with these SUSY particles in the external states. 
A number of studies have already addressed this issue, 
for production as well as for decay processes.
For squark and gluino production in hadron collisions,
the  NLO QCD corrections 
are available~\cite{Beenakker:1997ch}; 
for squark-pair production
in $e^+e^-$ collisions, the NLO QCD 
are also known, together with the Yukawa corrections~\cite{Eberl:1996wa}. 
Concerning the subsequent
squark decays into charginos/neutralinos, the QCD corrections were
presented in~\cite{Kraml:1996kz,Djouadi:1997wt}\footnote{The gluino
decay  channel, which can be overwhelming for $m_{\tilde
  q}>m_q+m_{\tilde g}$, was studied in~\cite{Beenakker:1996de}. Here we
will assume $m_{\tilde g}>m_{\{\tilde t_a,\tilde b_a\}}$.},
whereas the Yukawa corrections were given in~\cite{Guasch:1998as}.
In this last work large corrections were found. 
They were derived, however, 
in the \textit{higgsino} approximation for the chargino; 
hence, a full computation is required 
to consolidate the significance of the loop effects.

In this letter we present, for the first time, 
a complete
one-loop computation of the electroweak
radiative corrections to the 
partial decay widths of squarks 
into quarks and charginos/neutralinos, 
\begin{equation}
\Gamma(\tilde{q} \to q' \chi)\,\,,
\label{eq:gammadef}
\end{equation}
with $\tilde{q}=\sbottom,\,\stopp$, 
$q=b,\,t$,
$\chi=\neut_{\alpha},\,\chi^\pm_i$. 
We present the basic structure of the corrections, 
and illustrate their main features and their significance 
in representative numerical examples. 
The fine details, as well as a comprehensive analysis, 
will be presented elsewhere~\cite{GHS2}. 
Although the explicit results, for definiteness,
are displayed for the third squark generation, 
our analytic results are valid for all kinds of
squarks. 

In processes with exclusively SM particles in external states,
it is possible to divide the  1-loop contributions into
SM-like and non-SM-like subclasses.
This separate treatment is often used in the literature, 
and it is useful since it allows to make the
computation in small steps, checking each sector individually.
As a distinctive feature of the radiative corrections to processes with
supersymmetric particles in the external legs, this separability is lost.
In this kind of processes the ultraviolet (UV)
divergences of diagrams with virtual SM particles cancel the
UV divergences of diagrams with non-SM particles. 
Any partial computation would yield
UV-divergent and thus meaningless results.
For this reason
we have to compute all the SM and non-SM loop contributions
at once, with the proper counter terms involving
the renormalization of almost the full MSSM.

We have used an on-shell renormalization scheme, which can be obtained 
by extending 
the on-shell scheme of~\cite{Bohm:1986rj} 
to the MSSM. For the SM sector, the gauge-boson and fermion masses
are treated as input parameters; the electron charge is defined
in the Thomson limit, and the weak mixing angle is given in terms of
$\sw^2\equiv \swp^2\equiv1-\mws/\mzs$. 
The SM sector and its 
renormalization is described in~\cite{Bohm:1986rj} 
and will not be  repeated here.
On the other hand, different conventions for the SUSY sector exist; 
therefore, a few comments on our procedure are in order. Concerning
the Higgs sector, one mass has to be specified as an input quantity
(for which we take the mass of the $H^\pm$), and a definition
for \tb, the ratio between the vacuum expectation value of the two
Higgs-boson doublets, $v_2/v_1$, is needed.
Following~\cite{Dabelstein:1995hb}, an indirect definition of \tb\  
is given by the requirement\footnote{For possible other definitions see
e.g.~\cite{Coarasa:1996qa}.} 
$\delta v_2/v_2={\delta v_1}/{v_1}$.
For the sfermion sector, we use the input and the renormalization conditions
as described in~\cite{Guasch:1998as}, which fixes the squark masses 
and the squark-mixing angles~$\theta_{\sq}$.

Finally, as a new ingredient, we have to
specify the chargino--neutralino sector.  
The tree-level
mass matrices are well known, but we list  them in order to settle
the conventions:
\begin{eqnarray}
{\cal M}&=&\pmatrix{
M&\sqrt{2} M_W \sin{\beta} \cr
\sqrt{2} M_W \cos{\beta}&\mu
}\,\,,\nonumber\\
{\cal M}^0&=&
\pmatrix{
M^\prime&0&M_Z\cos{\beta}\swp&-M_Z\sin{\beta}\swp \cr
0&M&-M_Z\cos{\beta}\cwp&M_Z\sin{\beta}\cwp\cr
M_Z\cos{\beta}\swp&-M_Z\cos{\beta}\cwp&0&-\mu \cr
-M_Z\sin{\beta}\swp&M_Z\sin{\beta}\cwp&-\mu&0
}\,\,\,.
\label{eq:mmassacplus}\label{eq:mmassaneut}\end{eqnarray}
$M$ and $M'$ are the $SU(2)_L$ and $U(1)_Y$ soft-SUSY-breaking gaugino
masses. These mass matrices are diagonalized by 
unitary matrices $U,V,N$ via 
\begin{equation}
\begin{array}{lcccl}
U^* {\cal M} V^\dagger&=&{\cal M}_D&=&{\rm diag}\left(M_1,M_2\right)\,\,(0<M_1<M_2)\,\,,\\
 N^*{\cal M}^0 N^\dagger &=&{\cal M}^0_D&=&
{\rm diag}\left(M_1^0,M_2^0,M_3^0,M_4^0\right)\,\,(0<M_1^0<M_2^0<M_3^0<M_4^0)\,\,.
\end{array}\label{eq:defUVN}
\end{equation}
This sector contains six particle masses, but only 
three free parameters are available for an independent renormalization.
As a consequence, we are not allowed  to impose
on-shell conditions for all the particle masses.
For the independent input parameters, we choose:
the masses of the two charginos and the mass of the lightest neutralino. 
By introducing counterterms for all the independent parameters
in~(\ref{eq:mmassaneut}), we are able to relate the counterterms of the
fundamental parameters to the mass-counterterms $\delta M_i$ of the
charginos. Similarly to~\cite{Pierce:1994gj} 
we find:
\begin{eqnarray}
M_1 \,\delta M_1 + M_2 \,\delta M_2 &=&  M \,\delta M + \mu \,\delta\mu+ \delta
\mws , \nonumber\\
M_1 M_2
\left(M_1 \,\delta M_2+M_2\,\delta M_1\right)&=&\left(M \mu -\mws
  \sin(2\beta)\right)\bigg[M \,\delta \mu + \mu \,\delta M\nonumber\\
&&-\mws \,\delta\sin(2\beta)-\sin(2\beta) \,\delta\mws\bigg] .
  \label{eq:defcountcha} 
\end{eqnarray}
The mass counterterms
$\delta M_i$ are fixed using the on-shell scheme
relation, in the convention 
of~\cite{Bohm:1986rj} (but with opposite sign for $\Sigma$), 
\begin{equation}
  \delta M_i=-\frac{1}{2}\left(\Sigma_L^i(M_i^2)+\Sigma_R^i(M_i^2)\right)-\Sigma_S^i(M_i^2)\,\,,
  \label{eq:defdeltami}  
\end{equation}
where $\Sigma_{\{L,R,S\}}^i$ denote
the one-loop unrenormalized left-, right-handed and scalar components of
the self-energy for the $i$th-chargino. 
$\delta M'$ is determined from  the lightest neutralino
mass, inverting the relation
\begin{equation}
  \label{eq:defcountneut}
N^*_{1 \alpha} \delta{\cal M}^0_{\alpha\beta} N_{ 1\beta}^{*} = \delta M_1^0\,\,,
\end{equation}
where the neutralino-mass counterterm 
$\delta M_1^0$ is fixed by the on-shell condition for $\chi^0_1$,
in analogy to~(\ref{eq:defdeltami}). It is a
non-trivial check that with the counterterms determined  in
eqs.~(\ref{eq:defcountcha}) and (\ref{eq:defcountneut}), the 
one-loop masses for the residual neutralinos,
computed as the pole masses, are UV-finite.
The one-loop on-shell neutralino masses read
\begin{equation}
  M_\alpha^{0 \ {\rm os}}=M_\alpha^0+N^*_{\alpha\beta} \delta{\cal M}^0_{\beta\gamma} N_{\alpha\gamma}^{*}
  +\frac{1}{2}\left(\Sigma_L^\alpha(M_\alpha^{02})+\Sigma_L^\alpha(M_\alpha^{02})\right)+\Sigma_S^\alpha(M_\alpha^{02})\,\, ,
  \label{eq:defmass1l}  
\end{equation}
where now the parameters of eq.~(\ref{eq:mmassaneut}) and the masses and
mixing matrices computed in~(\ref{eq:defUVN}) have to be regarded as
\textit{renormalized} quantities.

The choice of the lightest neutralino to fix the counterterm 
$\delta M'$ in~(\ref{eq:defcountneut}) is only efficient if it has
a substantial \textit{bino} component. If $M'\gg(|\mu|,M)$ then
$|N_{11}|\ll1$, and the extraction of $\delta M'$
from~(\ref{eq:defcountneut}) would amplify the radiative corrections
artificially. In this case
it would be better to extract $\delta M'$ from the
$\alpha$th neutralino, such that $|N_{1\alpha}|$ is large. 
This is, however, not relevant for the scenarios which are discussed
in this letter.
Notice also that our renormalization procedure 
makes use of positive-definite mass eigenvalues for charginos and
neutralinos, which require the introduction of some purely-imaginary 
non-zero elements in the $N$-matrix~(\ref{eq:defUVN}). Had we
chosen a 
real $N$-matrix, with some negative eigenvalues, the various
renormalization conditions would be plagued with the explicit sign of the
corresponding eigenvalue (see e.g.~\cite{Guasch:1997dk}).

At one-loop, also
mixing self-energies between the different neutralinos and charginos are
generated, which we write as follows:
\begin{equation}
  -i \hat\Sigma^{\alpha\beta}(k^2)=-i \left(\hat \Sigma^{\alpha\beta}_L(k^2)
   \slas{k} \PL + \hat \Sigma^{\alpha\beta}_R(k^2)   \slas{k} \PR
   +\hat\Sigma^{\alpha\beta}_{SL}(k^2) \PL + \hat\Sigma^{\alpha\beta}_{SR}(k^2) \PR\right)\,\,,\,\,\alpha\neq\beta\,\,,
  \label{eq:mixingneut}
\end{equation}
with
$\hat\Sigma$ denoting the renormalized two-point functions, 
and with the chirality projectors $P_{\{L,R\}}=1\mp\gamma_5$. 
For the neutralinos,
the renormalized self-energies~(\ref{eq:mixingneut}) are related to the
unrenormalized ones according to
\begin{equation}
  \hat\Sigma^{\alpha\beta}_{\{L,R\}}=\Sigma^{\alpha\beta}_{\{L,R\}}\,\,,\,\,
  \hat\Sigma_{SL}^{\alpha\beta}=\Sigma_{SL}^{\alpha\beta}{
    -N_{\alpha\gamma}\delta{\cal M}^{0*}_{\gamma\lambda} N_{\beta\lambda}}\,\,,\,\,
  \hat\Sigma_{SR}^{\alpha\beta}=\Sigma_{SR}^{\alpha\beta}{
    -N^*_{\alpha\gamma} \delta{\cal M}^{0}_{\gamma\lambda} N_{\beta\lambda}^{*}}\,\,,
  \label{eq:mixinneutren}  
\end{equation}
and analogous expressions hold for the charginos.

As far as vertex renormalization is concerned, the vertex counterterms are
already determined by the renormalization procedure described above.
In addition to the parameter renormalization,
we have introduced also  
a field-renormalization constant 
for each left- and right-handed chargino and neutralino field.
As an explicit example, we list
the renormalized bottom-sbottom-neutralino vertex. 
The tree-level interaction Lagrangian
reads~\cite{Coarasa:1996qa}\footnote{Note that our convention for the
  neutralino mass-matrices~(\ref{eq:mmassaneut}) is different from that of~\cite{Coarasa:1996qa}.}
\begin{equation}
\begin{array}{l}
  {\cal L}_{\neut b\tilde{b}}=
  +\displaystyle{\frac{g}{\sqrt{2}}}
  \sum_{a=1,2}\sum_{\alpha=1,\ldots ,4}
    \sba^*\nba\left(\Apab\PL+\Amab\PR\right)\,b
    +\mbox{\rm h.c.}\ ,\\ 
        \Apab = \Robc\left(\Nt^*-\frac{1}{3}\twp\No^*\right)
        -\sqrt{2}\lb \Rtbc\Nth^*\, ,\ 
        \Amab = -\sqrt{2}\lb \Robc\Nth
        -\frac{2}{3}\twp\Rtbc\No\,\,,
  \label{eq:bsbntree}  
\end{array}
\end{equation}
where the bottom-quark Yukawa coupling is 
$\lb=\mb/(\sqrt{2}\mw\cos\beta)$. 
Introducing the one-loop counterterms analogously
to~\cite{Guasch:1998as} we obtain the following counterterm
Lagrangian~\cite{Guasch:1998as,Coarasa:1996qa}
\begin{eqnarray}
  \delta {\cal L}_{\neut_\alpha b\sbottom_a}&\equiv&  \displaystyle{\frac{g}{\sqrt{2}}}
    \sba^*\nba\left(\delta\LApab\PL+\delta\LAmab\PR\right)\,b
    +\mbox{\rm h.c.}\nonumber\\
&=&   \frac{1}{2}\left\{
  \left[\dalpha+\frac{\cwp^2}{\swp^2} \left(\dmws-\dmzs\right)\right]
    + \dz_{a}\right\}  {\cal L}_{\neut_\alpha b\sbottom_a} 
  +{\cal L}_{\neut_\alpha
    b\sbottom_c}\dz_{ca}\nonumber\\
  &&+\Bigg\{\displaystyle{\frac{g}{\sqrt{2}}}\sba^*\nba\left[\Apab
  \frac{1}{2} \left(\Dzrneut{}_\alpha+\DZlbot\right) \PL
  +\Amab \frac{1}{2} \left(\Dzlneut{}_\alpha+\DZrbot\right)\PR\right]\,b\nonumber\\
  &&+\displaystyle{\frac{g}{\sqrt{2}}}\sba^*\nba\left(\delta\Apab
    \PL+\delta \Amab \PR\right)\,b
  +\mbox{ h.c.}\Bigg\}\,\,,\,\,(c\neq a)\,\,,\nonumber \\
  \delta \Apab&=&    -\sqrt{2} \left( \Drb_{2a} \lb N_{\alpha 3}
  +   \delta \lb R^{(b)}_{2a} N_{\alpha 3} \right)+
  \Drb_{1a}  \left(N_{\alpha 2}
  -\frac{1}{3} \twp N_{\alpha 1} \right) - \frac{1}{3}  \delta \twp R^{(b)}_{1a}
  N_{\alpha 1}
  \,\,,\nonumber\\
  \delta \Amab&=&-\sqrt{2} \left(\Drb_{1a} \lb N^*_{\alpha3}
  + \delta \lb R^{(b)}_{1a} N^*_{\alpha3} \right)
  -\frac{2}{3}  \twp N^*_{\alpha1}  \left(
  \dtow R^{(b)}_{2a}+\Drb_{2a}         
   \right)\,\,,\nonumber \\
  \frac{\delta \lb}{\lb}&=&\dmb-\frac{1}{2}
  \dmws-\Dcosb\,\,,\,\,
  \delta{t_W}=\frac{1}{c_W^2}\left(\frac{\delta
    M_Z^2}{M_Z^2}-\frac{\delta M_W^2}{M_W^2} \right) ,
  \label{eq:counterl}
\end{eqnarray}
where $\delta\alpha$, $\delta M_{W,Z}^2$ are the charge and mass counterterms
for the MSSM, as given
in~\cite{Garcia:1994sb}. 
The renormalized one-loop part of the amplitude for the decay
$\sbottom_a \rightarrow b \chi^0_\alpha$ 
can then be written as
\begin{equation}
  -i T_{a\alpha}^{\rm loop}=-i
   \frac{g}{\sqrt{2}}\bar{u}_b \left[C_{+\alpha\beta}\PR
     +C_{-\alpha\beta}\PL\right] v_{\neut_\alpha}\,\,;\,\,
   C_{\pm\alpha\beta}=\delta\LApmabc+\LApmab{}^{\Sigma}
   +\LApmab{}^{\rm 1PI}\,\,.
  \label{eq:renormamp} 
\end{equation}
Besides the counterterms $\delta\Lambda$ from eq.~(\ref{eq:counterl}),
it contains the one-loop contribution $\Lambda^{\rm 1PI}$ 
to the one-particle-irreducible three-point vertex function
and the quantity
$\Lambda^{\Sigma}$  
representing the contribution of the neutralino-mixing
self-energies~(\ref{eq:mixingneut}),  
\begin{equation}
 \LApmab{}^{\Sigma}=\sum_\beta \Apmbbc
         \frac{
                 M_{\alpha}^0 \hat\Sigma_{SL}^{\beta\alpha}
                 +M_{\beta}^0 \hat\Sigma_{SR}^{\beta\alpha}
                 +M_{\alpha}^0 M_{\beta}^0 \hat\Sigma_L^{\beta\alpha}
                 +M_{\beta}^0{}^2 \hat\Sigma_R^{\beta\alpha}
                 }{M_{\alpha}^0{}^2-M_{\beta}^0{}^2}\,\, .  
  \label{eq:mixingeffect}
\end{equation}
Due to the presence of photon loops, the amplitude~(\ref{eq:renormamp})
is infra-red divergent;
hence, bremsstrahlung of real photons has to be added  
to cancel this divergence. We therefore
include in our results the radiative partial
decay width $\Gamma(\sbottom_a \to b\neut_\alpha \gamma)$, 
including both the soft and the hard photon part. 
So finally, the complete one-loop electroweak correction is given by
\begin{equation}
\begin{array}{rcl}
\delta^{a\alpha}_0&=&
\frac{\Gamma(\sbottom_a\to b\neut_\alpha)}{\Gamma^0(\sbottom_a\to b\neut_\alpha)}-1
= \delta^{a\alpha}_{0 {\rm virt}}+\frac{\Gamma(\sbottom_a \to b\neut_\alpha
  \gamma)}{\Gamma^0(\sbottom_a\to b\neut_\alpha)}\,\,,\\[0.5cm]
\delta^{a\alpha}_{0 {\rm virt}}&=&
2{\rm Re}\left( \frac{ (\msbas-M_\alpha^{02}-\mbs) (\Apab \Cpab + \Amab
    \Cmab) - 2 M_\alpha^0 \mb (\Apab \Cmab+\Amab \Cpab)}{(\msbas-M_\alpha^{02}-\mbs) (|\Apab|^2+|\Amab|^2) -4 M_\alpha^0 \mb {\rm Re}(\Apab \Amabc)}\right)\,\,,
  \label{eq:deltadef}
\end{array}
\end{equation}
for the neutralino decay channels,
and by corresponding expressions $\delta^{ai}_+$ for the chargino
channels, as well as for the top-squark decays.

The loop computation itself is rather tedious, since there is a huge
number of diagrams to compute. This is better done by
means of automatized tools.
The computation of the loop diagrams has been done by using the Computer
Algebra Systems \textit{FeynArts} and
\textit{FormCalc}~\cite{Kublbeck:1990xc,Hahn:1998yk}. 
We have
produced a set of Computer Algebra programs that compute the one-loop
diagrams (and the bremsstrahlung corrections), which are then plugged into
a \textit{Fortran} code for the numerical evaluation with the help of the
one-loop routines
\textit{LoopTools}~\cite{Hahn:1998yk}. 
A number of checks has been made  on the results. 
The UV and infra-red
finiteness of the result, relying on the relations between the
different sectors of the model, is a non-trivial check. We also have
recovered results already available in the
literature; for instance, we used our set of programs to reproduce the strong
corrections of~\cite{Djouadi:1997wt}, and, using the \textit{higgsino}
approximation, we could also reproduce the results
of~\cite{Guasch:1998as}. Moreover we also checked that, when using the
$\overline{\mbox{MS}}$-scheme, the one-loop corrections to neutralino
and chargino masses reproduce those of~\cite{Pierce:1994gj}. 

Although we consider the chargino and neutralino masses as input
parameters, in our numerical study we treat them in a slightly different
way. We choose a set of renormalized input parameters $(M,M',\mu)$, and
apply~(\ref{eq:mmassaneut}), (\ref{eq:defUVN}) to obtain the one-loop
renormalized masses. 
Of course, if SUSY would be discovered the procedure will
be the other way around, that is, the MSSM parameters will be computed
from the various observables measured, for example, from the chargino
production cross-section and asymmetries at 
the LC~\cite{Kneur:1999gy}. 
For a consistent treatment, the one-loop expressions for
these observables will have to be used~\cite{Blank:2000uc}.

\figsbottomone

As for the numerical analysis, we use a set of parameters relevant for
the next generation of colliders. The squark masses have been chosen in
the range of $\sim 300\GeV$. If the squarks have a mass around this scale,
they will be produced at significant rates not only at the LHC, but also at 
a LC with  $800\GeV$ center-of-mass energy. As for the gaugino sector, we
make use of the GUT relation between the gaugino soft-SUSY-breaking
mass parameters, $M'=5/3 \tan^2\theta_W M$.
The input parameters for the SUSY-electroweak sector are chosen to be
\begin{equation}
 \tb=  4\,\,,\,\,
 \mu    =  -100\GeV\,\,,\,\,
 M       =  150\GeV\,\,,\,\,
 \mHp=120\GeV\,\, .
\end{equation}
For the quark-squark sector we take
\begin{eqnarray} 
&&\mb=5 \GeV\,\,,\,\,
m_{\sbottom_1}=300\GeV\,\,,\,\,m_{\sbottom_2}=m_{\sbottom_1}+5\GeV\,\,,\,\,\theta_{\sbottom}=0.3\,\,,\nonumber\\
&&\mt=175 \GeV\,\,,\,\, m_{\stopp_1}= 300\GeV\,\,, \,\,
 \theta_{\stopp}=0.6\,\,,
\end{eqnarray}
the rest of squarks are given a mass of $\sim 1 \TeV$ with a mixing
angle $\theta_{\tilde q}=\pi/4$. All over our numerical results we
apply the (approximate) condition for the non-existence of
colour-breaking minima, demanding that the (computed) values of the
trilinear soft-SUSY-breaking couplings $A_q$ do not exceed 
$3 m_{\tilde q}$~\cite{Frere:1983ag}.

In Fig.~\ref{fig:sbottom} we present the results on the lightest-sbottom 
decay. Fig.~\ref{fig:sbottom}a presents the tree-level
prediction for the various branching ratios ($BR$) as a function of the
lightest-sbottom mass. We see that, aside from the
third neutralino channel, all the channels have an appreciable
branching ratio whenever they are possible. The opening of the bosonic
decay channel ($\sbottom_1\to \stopp_1 W^-$) is clearly visible
at $m_{\sbottom_1}=m_{\stopp_1}+\mw=380\GeV$. The charged Higgs boson
channel ($\sbottom_1\to \stopp_1 H^-$) opens at
$m_{\sbottom_1}=m_{\stopp_1}+\mHp=420\GeV$, but its partial width is
much smaller (for the chosen value of the parameters), and hence we can
not visualize its effect in Fig.~\ref{fig:sbottom}a. Whenever the
bosonic decay channels are open they amount to a large fraction of the
branching ratio~\cite{Bartl:1998xk}, and they suffer from large
radiative corrections~\cite{Bartl:1997pb}. 
In Fig.~\ref{fig:sbottom}b
and c we present the one-loop electroweak
radiative corrections to the chargino
and neutralino channels, respectively. The lightest-neutralino channel is
specially important, in that it is always open when we require the
lightest neutralino to be the Lightest Supersymmetric Particle. When
making a numerical scan over the MSSM parameters the corrections show a
rich structure, owing to the large number of thresholds,
pseudo-thresholds, etc.  For example, the variation of the
higgsino-mass parameter $\mu$ does not only change the chargino--neutralino
masses and mixing angles, but also the value of the 
soft-SUSY-breaking trilinear squark couplings, and the mass of the
heaviest top-squark. We see in Fig.~\ref{fig:sbottom}a the
opening of the $\sbottom_1\to t \cmin_2$ channel at 
$m_{\sbottom_1}=365\GeV$. This threshold is accompanied by a
corresponding divergence of the one-loop corrections to the 
lightest-chargino 
channel ($\delta^{11}_{+}$ in Fig.~\ref{fig:sbottom}b), and
also in the various neutralino channels (Fig.~\ref{fig:sbottom}c). Of
course, the (divergent) corrections near the threshold do not have
a physical meaning, since perturbation theory breaks down. In the
\textit{light-sbottom} region ($m_{\sbottom_1}\lsim 350 \GeV$) the
branching ratio is neutralino-dominated. In this region the corrections 
amount to a 5-10\% positive corrections for all of the channels. After
the opening of the $\cmin_2$ channel the picture changes a little. The
third neutralino can get corrections up to 30\%, its $BR$, however, is
smaller that 1\%; these large corrections are thus of little interest. The
rest of the channels continue with moderate corrections of the order
of 5-10\%; a special address to the heaviest-neutralino channel
($\neut_4$) with a 15\% correction and a non-negligible $BR\simeq1-8\%$
is in order.
At this point, however, we do not know yet the net effect on the
corrected branching ratios
since the electroweak corrections are similar in all decay channels.
For quantitative statements
also the QCD corrections~\cite{Kraml:1996kz,Djouadi:1997wt} 
have to be included.

\figsbottomtwo

The corresponding tree-level branching ratios and radiative corrections for the
heaviest bottom-squark are displayed in Fig.~\ref{fig:sbottom2}. They show
a similar pattern to that of the lightest sbottom, but with more
relevance for the chargino channels. We note that in this case negative
radiative corrections are attained for the lightest neutralino
(Fig.~\ref{fig:sbottom2}c). The maximum radiative corrections to the
neutralino channel are 15\% (for the heaviest neutralino), which has a
non-negligible branching ratio all over the allowed sbottom-mass
range. Notice that the very different corrections to the chargino
channels ($\sim 17\%$ for $\cmin_1$ and $\sim 4\%$ for $\cmin_2$) at
large sbottom masses will translate into corresponding corrections to
the branching ratios. 

\figstop
 
We now turn our attention to the lightest-stop decay channels. In
Fig.~\ref{fig:stop} we present the branching ratios and the corrections
to all possible decays as a function of the lightest top-squark
mass. The chargino decay channels are important whenever
one of the charginos is lighter than the top-squark. The
top-squark is expected to be one of the lightest squarks of the model
(due to large squark mixing), it could be even lighter than the top
quark. In this latter case, if the chargino channel is closed, it would
decay through Flavour Changing Neutral Currents, 
$\stopp_1\to c\neut_1$~\cite{Hikasa:1987db}. These two decay channels
have been used for a experimental simulation to extract with high
precision the top-squark parameters at the LC~\cite{Berggren:1999ss}.
For the chosen set of parameters the neutralino channels have never a
$BR$ larger that $20\%$. When the lightest top-squark is lighter than
$\sim200\GeV$ the only possible decay channel has corrections to the
width up to $20\%$ (Fig.~\ref{fig:stop}b). These are, however, of little
practical interest, since the $BR$ is 100\% in any case.
Above this mass, both chargino channels
receive corrections in the 5-10\% range. The neutralino channels obtain
corrections in the 5-15\% range, but, as in the case of the sbottom
decays, the largest corrections occur in the channels with smallest
branching ratios. Again, several threshold structures are visible in the
plot, accompanied by the corresponding divergence of the corrections.

In summary, we have presented the set of full electroweak corrections to
squark decays into quarks and charginos/neutralinos. These corrections
can be sizeable, and therefore they have to be taken into account for the
extraction of the MSSM parameters from experiment. A sample of the
numerical results has been presented. The final impact on the branching
ratios will also depend on the strong
corrections~\cite{Kraml:1996kz,Djouadi:1997wt}. 
A combination of the two sets of
corrections, as well as a comprehensive analysis will be presented
elsewhere~\cite{GHS2}. 


\bigskip
\noindent\textbf{Acknowledgments:} \\
The calculations have been done using the QCM cluster of the 
DFG Forschergruppe ``Quantenfeldtheorie, Computeralgebra und
Monte-Carlo Simulation''.
We are thankful to T. Hahn for
his help regarding the Computer Algebra system, and  to
A. Vicini and D. St{\"o}ckinger for useful discussions.
The collaboration is part of the network ``Physics at Colliders'' of the
European Union under contract HPRN-CT-2000-00149.
The work of J.G. is supported by the 
European Union under contract No. HPMF-CT-1999-00150. The work of
J. S. has been supported in part by CICYT under project No. AEN99-0766.

\providecommand{\href}[2]{#2}\begingroup\raggedright
\endgroup

\end{document}